\documentclass[pra,aps]{revtex4-2}

\usepackage{graphicx}
\usepackage{dcolumn}
\usepackage{bm}

\usepackage{xcolor}
\usepackage{CJKutf8}
\begin{document}
\begin{CJK*}{UTF8}{gbsn}
\title{Full characterization of biphotons with a generalized quantum interferometer}

\author{Baihong Li(李百宏)$^{1}$}
\email{li-baihong@163.com}

\author{Changhua Chen(陈昌花)$^{1}$}
\author{Boxin Yuan(袁博欣)$^{1}$}
\author{Xiaofei Zhang(张晓斐)$^{1}$}



\author{\\ Ruifang Dong(董瑞芳)$^{2,3}$}%
 \email{dongruifang@ntsc.ac.cn}

\author{Shougang Zhang(张首刚)$^{2,3}$}%


\author{Rui-Bo Jin(金锐博)$^{4}$}%
 \email{jin@wit.edu.cn}

\affiliation{$^{1}$
School of Physics and Information Science, Shaanxi University of Science and Technology, Xi’an 710021, China
}%
\affiliation{$^{2}$
Key Laboratory of Time and Frequency Primary Standards, National Time Service Center, Chinese
Academy of Sciences, Xi’an 710600, China
}%

\affiliation{$^{3}$
School of Astronomy and Space Science, University of Chinese Academy of Sciences, Beijing 100049, China
}%

\affiliation{$^{4}$
 Hubei Key Laboratory of Optical Information and Pattern Recognition, Wuhan Institute of Technology,
Wuhan 430205, China
}%

\begin{abstract}
Entangled photons (biphotons) in the time-frequency degree of freedom play a crucial role in both foundational physics and advanced quantum technologies. Fully characterizing them poses a key scientific challenge. Here, we propose a theoretical approach to achieving the complete tomography of biphotons by introducing a frequency shift in one arm of the combination interferometer. Our method, a generalized combination interferometer, enables the reconstruction of the
full complex joint spectral amplitude associated with both frequency sum and difference in a single interferometer. In contrast, the generalized Hong-Ou-Mandel and N00N state interferometers only allow for the partial tomography of biphotons, either in frequency difference or frequency sum. This provides an alternative method for full characterization of an arbitrary two-photon state with exchange symmetry and holds potential for applications in high-dimensional quantum information processing.

\end{abstract}

\maketitle


\section{\label{sec:1}INTRODUCTION}

Entangled photon sources in the time-frequency degree of freedom play a crucial role in both foundational physics and advanced quantum technologies, such as high-dimensional quantum information processing \cite{Optica2017,NP2019,SCIS2023}. Since high-dimensional information can be naturally encoded in the time and frequency degrees of freedom, such entangled sources have the potential to improve the robustness and key rate of quantum communication protocols \cite{PRL2007,OE2013,NJP2015}, quantum enhanced sensing \cite{NP2011}, and achieve more efficient and error-tolerant quantum computation \cite{NP2009}. These applications require well-characterized sources, and how to fully characterize them is a key scientific challenge.

A direct way to characterize the spectrum of entangled photon pairs (biphotons) is to measure the joint spectral intensity (JSI), which gives the probability of detecting the photons with given frequencies \cite{JMO2018}, and the joint temporal intensity (JTI), which gives the probability of detecting the photons at given arrival times \cite{PRL2008,PRL2018}. Joint measurements of JSI and JTI in both frequency and time have enabled partial characterization of entangled ultrafast photon pairs, but they are still unable to provide the full two-photon
state \cite{PRL2018}. An indirect way is quantum Fourier-transform spectroscopy established by the extended Wiener–Khinchin theorem \cite{JIN2018,PRA2022,PRApplied2022}, where the spectral information of entangled light can be extracted by performing a Fourier transform on its time-domain interferograms obtained from the Hong-Ou-Mandel (HOM) interferometer \cite{PRL1987,RPP2021}, the N00N state interferometer \cite{NOON,NOON1,NOON2} (also called the Mach-Zehnder interferometer), or their combination \cite{LiPRA2023}.

However, both of these methods are insensitive to the phase and therefore cannot reveal the phase information of biphotons \cite{PRL2018,PRApplied2018}. To address this issue, some efforts have been made based on phase retrieval algorithms that are widely used in classical ultrafast optics \cite{PRA2019,PRR2019}. Recently, Davis et al. \cite{Optica2020} demonstrated a technique for determining the full quantum state of biphotons using electro-optic shearing interferometry. Some relevant review articles can be found in Ref. \cite{AVS2020,AQT2021}.

Another way to acquire the phase information of biphotons is to introduce a frequency shift in one arm of interferometers, which has been demonstrated in a conjugate-Franson interferometer for the time-energy-entangled resource \cite{Chen2021RPL}. Later, N. Fabre \cite{Fabre2022} theoretically proposed a generalized HOM interferometer that allows for the reconstruction of the amplitude and phase of the joint spectral amplitude (JSA) associated with frequency difference for any symmetric JSA. Additionally, a generalized N00N state interferometer was introduced, allowing the reconstruction of the amplitude and phase of the JSA associated with frequency sum for symmetric JSA and frequency difference for antisymmetric JSA. However, it is impossible to obtain the full complex JSA associated with both frequency difference and sum using either of these two interferometers.

In this paper, we expand the combination interferometer that we recently proposed in \cite{LiPRA2023} and develop a generalized version by introducing a frequency shift in one arm of the interferometer. It is found that with the generalized combination interferometer, it is possible to reconstruct the full complex JSA associated with both frequency sum and difference for symmetric JSA or antisymmetric JSA within a single interferometer. This allows us to perform the full tomography of biphotons. In contrast, the generalized HOM and N00N state interferometers only enable partial tomography of biphotons, either in frequency difference or frequency sum. Additionally, we discuss an experimental feasibility involving biphoton sources with different symmetries, frequency shifts, and postprocessing of the coincidence data as well as the possible experimental difficulties with the proposed approach.

The rest of the paper is organized as follows. In Sec. \ref{sec:2}, we describe the generalized HOM interferometer and discuss the possibility of the reconstruction of the
full complex JSA associated with the frequency difference for any symmetric JSA. At the end of this part, as an example,  we analyse the phase sensitivity of such a generalized interferometer and show the simulated results for a Gaussian input state without and with a quadratic spectral phase. In Sec. \ref{sec:3}, we describe the generalized N00N interferometer and discuss the possibility of the reconstruction of the
full complex JSA associated with frequency sum for symmetric JSA and frequency difference for antisymmetric JSA. The simulated results are also shown for a Gaussian input state without and with a quadratic spectral phase. In Sec. \ref{sec:4}, we describe the generalized combination interferometer and discuss the possibility of the reconstruction of the
full complex JSA associated with both the frequency difference and sum for symmetric JSA or antisymmetric JSA. Section \ref{sec:5} compares the results obtained with three generalized interferometers and discusses the experimental feasibility and possible difficulties of performing the full tomography of biphotons with these generalized interferometers. Section \ref{sec:conclude} summarizes the results and concludes the paper.

\section{\label{sec:2}Generalized HOM interferometer}

\begin{figure}[th]
\begin{picture}(400,310)
\put(0,0){\makebox(405,300){
\scalebox{0.8}[0.8]{
\includegraphics{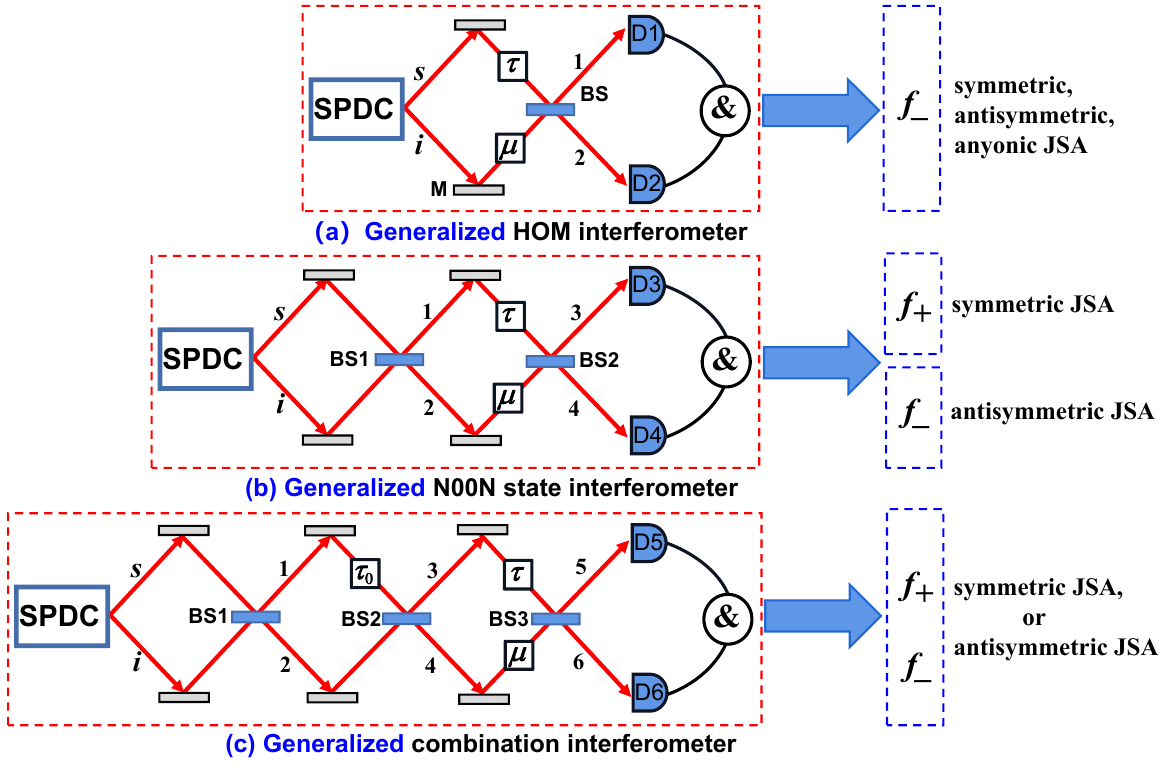}
}}}
\end{picture}
\caption{\label{Fig1}
(a) Generalized HOM interferometer, allows for the reconstruction of the amplitude and phase of $f_-$ for any symmetric JSA. (b) Generalized N00N state interferometer, allows for the reconstruction of the amplitude and phase of $f_+$ for symmetric JSA and of $f_-$ for antisymmetric JSA. (c) Generalized combination interferometer,allows for the reconstruction of the amplitude and phase of both $f_+$ and $f_-$ for symmetric or antisymmetric JSA. $\tau$ represents the time delay between two arms of interferometers, and $\mu$ the frequency shift. If there is no frequency shift $\mu$ in the above interferometers, i.e., $\mu=0$, they would revert to the original ones and only allow for the reconstruction of the amplitude of the corresponding $f$, namely $|f|$. M:Mirror, BS(50/50): beamspliter, D: detector, \boldsymbol{\&} : coincidence measurement.}
\end{figure}

The setup of a generalized HOM interferometer is illustrated in Fig. \ref{Fig1}(a), where a frequency shift $\mu$ is introduced in the idler arm. Assuming that the biphotons are generated, for instance, by spontaneous parametric down-conversion (SPDC). The coincidence probability between two detectors D1 and D2, as functions of the time delay $\tau$ and the frequency shift $\mu$ for the generalized HOM interferometer, can be expressed as
\begin{equation}
\label{R-HOM}
R(\tau,\mu)=\frac{1}{4}\int_{0}^{\infty}\int_{0}^{\infty}d\omega_s d\omega_i\Big|f(\omega_i+\mu,\omega_s)e^{-i\omega_s \tau}-f(\omega_s,\omega_i+\mu)e^{-i\omega_i \tau}\Big|^2.
\end{equation}
where $f(\omega_s,\omega_i)$ represents the JSA of the signal(s) and idler(i) photons. In general, the JSA cannot be factorized as a product of functions $f(\omega_s)$ and $f(\omega_i)$, revealing a frequency entanglement between two photons with frequency $\omega_s$ and $\omega_i$. However, it can be expressed in terms of collective coordinate $\omega_+$ and $\omega_-$ and generally decomposed as follows \cite{PRApplied2018,Fabre2022},
\begin{equation}
\label{JSA}
f(\omega_s,\omega_i)=f_+(\omega_+)f_-(\omega_-)=|f_+(\omega_+)||f_-(\omega_-)|e^{i\phi_+(\omega_+)}e^{i\phi_-(\omega_-)}.
\end{equation}
where $\omega_{\pm}=(\omega_s\pm\omega_i)/2$ and $f_+$ can be used to model the energy conservation in the SPDC process, and it depends on the spectral profile of the pump light. $f_-$ is the phase-matching function, which can have various
forms depending on the considered nonlinear crystal and the
method to achieve phase matching. The terms $|f_{\pm}|$ and $\phi_{\pm}$ denote the amplitude and phase of the JSA, associated with the frequency sum $\omega_+$ and the frequency difference $\omega_-$, respectively. The phase $\phi_{+}$ can be introduced by controlling the pump spectral phase, for instance, through a quadratic
spectral phase by chirping the pump as shown in Ref.\cite{Optica2020}, and the phase $\phi_{-}$ can be introduced by shaping the signal
and idler spectra with a programmable
phase spectral filter as demonstrated in Ref.\cite{Chen2021RPL}. If one seeks to know all information about biphotons, the tomography of the full complex JSA must be performed, i.e., the tomography of both the amplitudes of $|f_{\pm}|$ and the phases of $\phi_{\pm}$. 

The specific expression of Eq.(\ref{R-HOM}) depends on the exchange symmetry of the JSA. It is obvious that the exchange between $\omega_s$ and $\omega_i$ does not affect their sum $\omega_+$, thus the exchange symmetry of $f_+$. However, it will affect their difference $\omega_-$, resulting in a change in the exchange symmetry of $f_-$, e.g., $f(\omega_i,\omega_s)=f_+(\omega_+)f_-(-\omega_-)=f_+(\omega_+)f_-(\omega_-)$ for symmetric JSA, and $f(\omega_i,\omega_s)=f_+(\omega_+)f_-(-\omega_-)=-f_+(\omega_+)f_-(\omega_-)$ for antisymmetric JSA. This means that only the phase-matching function associated with $f_-$ affects the exchange symmetry of the JSA.

As derived in Appendix A, Eq.(\ref{R-HOM}) can be further simplified as
\begin{equation}
\label{R-HOM-W}
R(\tau,\mu)=\frac{1}{2}\Big(1-W_-(\tau,\mu)\Big).
\end{equation}
where
\begin{eqnarray}
\label{W}
W_-(\tau,\mu)&&=\int d\omega_-f_-(\frac{\mu}{2}-\omega_-)f_-^*(\frac{\mu}{2}+\omega_-)e^{-i2\omega_- \tau}.
\end{eqnarray}
is the chronocyclic Wigner distribution associated with the phase matching function $f_-$ \cite{SR-2013,PRA-2015}. The chronocyclic Wigner distribution is a real distribution and is negative (an entanglement witness \cite{OL-2008}) when $R(\tau,\mu)>1/2$. As such, it is a quasi-probability distribution. The cut of the Wigner distribution at $\mu=0$, where there is no frequency shift, corresponds to the original HOM interference result. In this case, assuming the phase distribution is symmetric, i.e., $\phi_-(\omega_-)=\phi_-(-\omega_-)$, $W_-(\tau,0)$ is a function of the Fourier transform of the modulus squared $|f_-(\omega_-)|^2$, which is insensitive to the phase. This allows us to obtain only the amplitude of the JSA, as shown in Ref.\cite{LiPRA2023}.  However, for the generalized HOM interferometer ($\mu\neq0$), we can see from Eq.(\ref{W}) that $W_-(\tau,\mu)$ is sensitive to the phase due to the introduction of the frequency shift $\mu$ \cite{Chen2021RPL}. Therefore, it is possible to to extract the phase information of $f_-(\omega_-)$. Assuming that $f_-(0)\neq0$, the following reconstruction formula can be used,
\begin{equation}
\label{f-}
f^*_-(\mu)=\frac{1}{2\pi f_-(0)}\int W_-(\tau,\mu)e^{i\mu \tau}d\tau,
\end{equation}
to perform the full tomography of $f_-$. The detailed derivation for Eq.(\ref{f-}) can be found in Appendix B. In practice, $W_-(\tau,\mu)$ can be obtained from the coincidence probability measured with the generalized HOM interferometer at different frequency shifts $\mu$. It is important to note that the HOM interferometer ($\mu=0$ in Fig. \ref{Fig1}(a)) always depends on the frequency difference, regardless of whether the JSA is symmetric, antisymmetric, or anyonic \cite{JIN2018,Fabre2022,LiPRA2023}. Therefore, the reconstruction formula Eq.(\ref{f-}) is suitable for any symmetric JSA. This has been shown schematically in the right part of Fig.\ref{Fig1}(a). When $\mu=0$, Eq.(\ref{f-}) represents a direct integral of the Wigner distribution, which corresponds to a marginal distribution of $|f_-|^2$. In this case, it only allows for the reconstruction of the amplitude of the corresponding $f_-$, namely $|f_-|$. To validate the effectiveness of the reconstruction formula Eq.(\ref{f-}), we have provided the proof and an example in Appendix B. The simulations of chronocyclic Wigner distribution for a Gaussian input state without and with a quadratic spectral phase in the example in Appendix B are presented in Fig. \ref{Fig2}.

\begin{figure}[th]
\begin{picture}(400,160)
\put(0,0){\makebox(405,150){
\scalebox{0.53}[0.53]{
\includegraphics{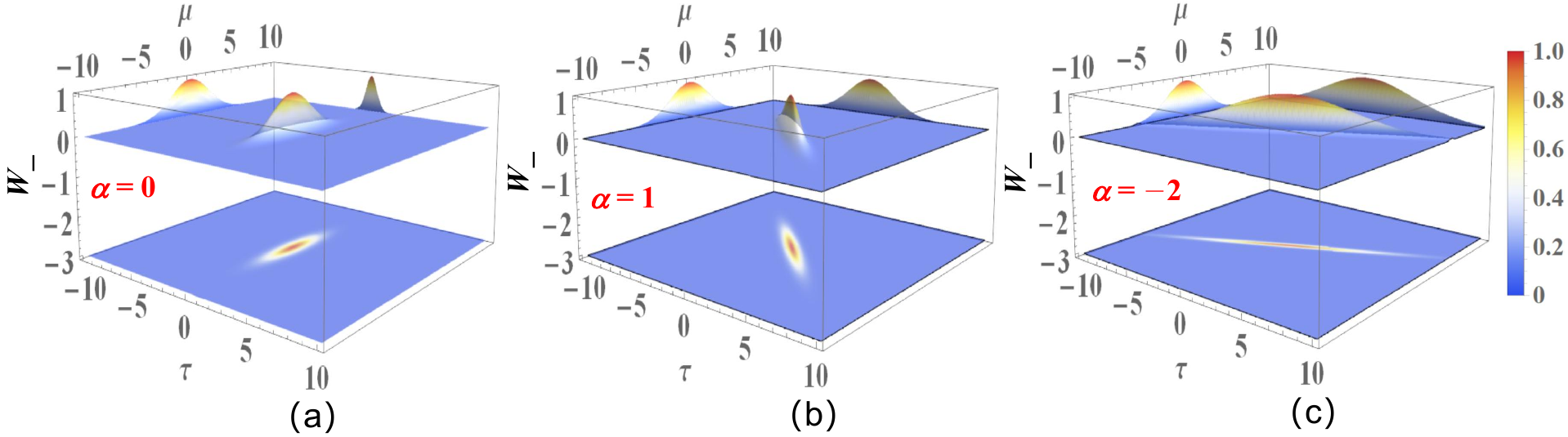}
}}}
\end{picture}
\caption{\label{Fig2}
Chronocyclic Wigner distribution as functions of the frequency shift $\mu$ and the time delay $\tau$ for a Gaussian input state are shown in (a) without a phase, i.e., $\alpha=0$, (b) with a quadratic spectral phase $\alpha=1$, and (c) with a quadratic spectral phase $\alpha=-2$. The quadratic spectral phase results in an increased time-frequency correlation and temporal variance.}
\end{figure}

\section{\label{sec:3}Generalized N00N state interferometer}
The setup of a generalized N00N state interferometer is illustrated in Fig. \ref{Fig1}(b), where a frequency shift $\mu$ is introduced in path 2. The coincidence probability between two detectors D3 and D4, as functions of the time delay $\tau$ and the frequency shift $\mu$ for the generalized N00N state interferometer, can be expressed as
\begin{equation}
\label{R-NOON}
R(\tau,\mu)=\frac{1}{2}\int \int d\omega_s d\omega_i\Big|f(\omega_s,\omega_i,\mu)(e^{-i(\omega_s+\mu) \tau}+1)(e^{-i(\omega_i+\mu) \tau}+1)+f(\omega_i,\omega_s,\mu)(e^{-i(\omega_i+\mu) \tau}-1)(e^{-i(\omega_s+\mu) \tau}-1)\Big|^2.
\end{equation}

Note that the N00N state is represented as $|2002\rangle=\frac{1}{\sqrt{2}}(|2,0\rangle+|0,2\rangle)$ in Fig. \ref{Fig1}(b), which has a photon number of 2. The specific form of Eq.(\ref{R-NOON}) depends on the exchange symmetry of the JSA.  If the JSA is symmetric, i.e., $f(\omega_s, \omega_i)=f(\omega_i, \omega_s)$, we can simplify Eq.(\ref{R-NOON}) as (see  Appendix C)
\begin{eqnarray}
\label{Rsys}
R_{S}(\tau,\mu)=\frac{1}{2}\Big(1+Re[F_+(\mu,\tau)]\Big).
\end{eqnarray}
where $F_+$ is the short-time Fourier transform (STFT) of
the function $f_+$ defined as
\begin{equation}
\label{F+}
F_+(\mu,\tau)=\int d\omega_+f_+(\omega_+)f_+^*(\omega_++\mu)e^{i2\omega_+ \tau}.
\end{equation}
If the JSA is antisymmetric, i.e., $f(\omega_s, \omega_i)=-f(\omega_i, \omega_s)$, we can further simplify Eq.(\ref{R-NOON}) as (see Appendix C)
\begin{eqnarray}
\label{Rant}
R_{A}(\tau,\mu)=\frac{1}{2}\Big(1+Re[e^{-i\mu \tau}F_-(\mu,\tau)]\Big).
\end{eqnarray}
where $F_-$ is the STFT of the function $f_-$ defined as
\begin{equation}
\label{F-}
F_-(\mu,\tau)=\int d\omega_-f_-(\omega_-)f_-^*(\omega_-+\mu)e^{-i2\omega_- \tau}.
\end{equation}
The additional phase factor $e^{-i\mu \tau}$ in Eq.(\ref{Rant}) stems from the symmetric time-frequency
displacement operator $\hat{D}(\mu,\tau)$ \cite{Fabre2022},
\begin{equation}
\label{D}
\chi_{\psi}(\mu,\tau)=\langle\psi|\hat{D}(\mu,\tau)|\psi\rangle=e^{-i\mu \tau/2}\int d\omega_- f_-(\omega_--\mu)f_-^*(\omega_-)e^{i\omega_- \tau}.
\end{equation}
where $\hat{D}(\mu,\tau)=e^{-i\mu \tau/2}\int d\omega e^{-i\omega \tau}|\omega+\mu\rangle\langle\omega|$ \cite{Fabre2020}.
\begin{figure}[th]
\begin{picture}(400,340)
\put(0,0){\makebox(405,330){
\scalebox{0.6}[0.6]{
\includegraphics{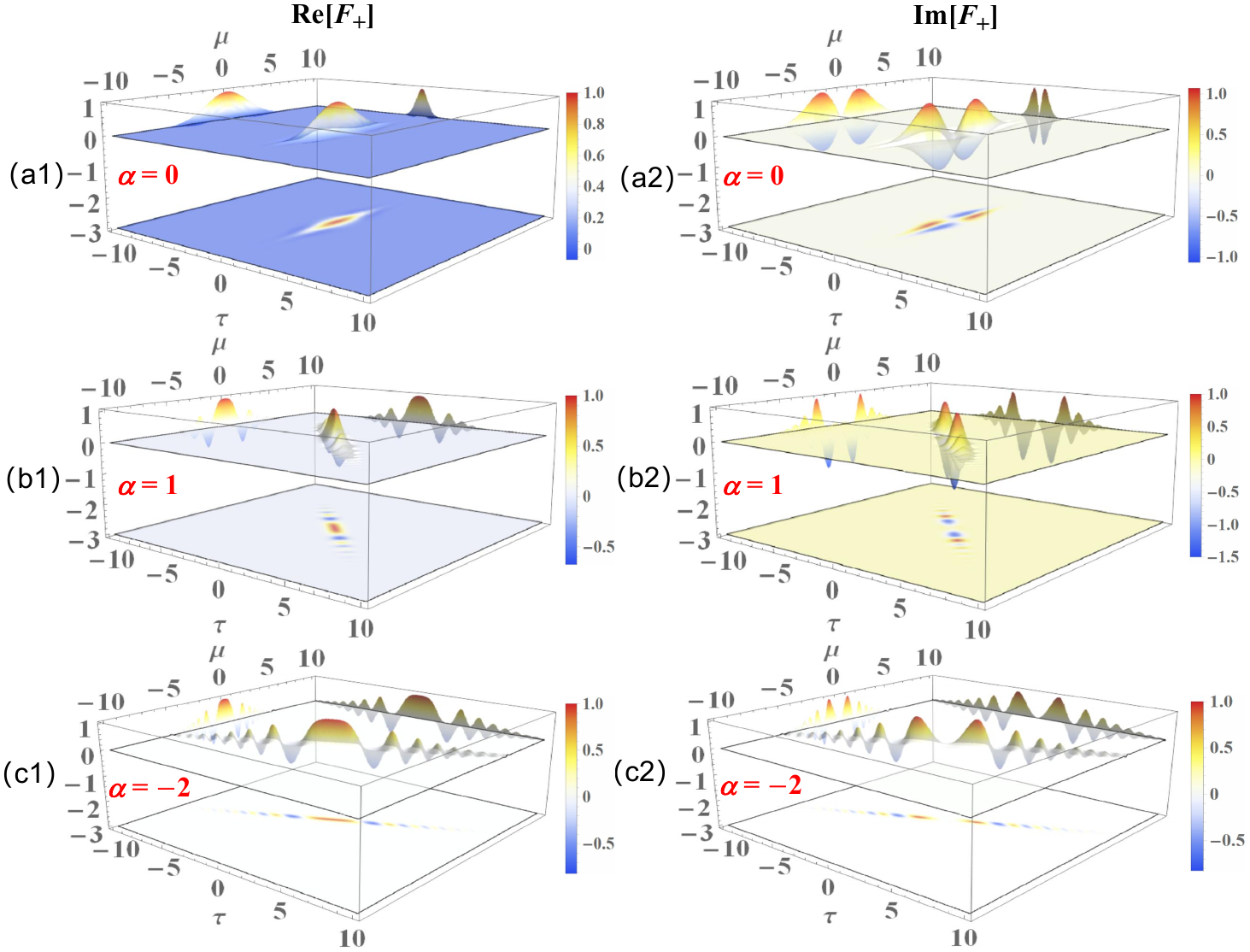}
}}}
\end{picture}
\caption{\label{Fig3}
The real (left) and imaginary parts (right) of $F_+$ as functions of the frequency shift $\mu$ and the time delay $\tau$ for a Gaussian input state are depicted as follows: (a) without a phase, i.e., $\alpha=0$, (b) with a quadratic spectral phase $\alpha=1$, and (c) with a quadratic spectral phase  $\alpha=-2$.}
\end{figure}

Note that $F_{\pm}$ is essentially the Fourier transform of $f_{\pm}(\omega)f_{\pm}^*(\omega+\mu)$, a complex function representing the phase and magnitude of the function $f_{\pm}$ with respect to time and frequency. If $\mu=0$, $F_{\pm}$ becomes a Fourier transform of $|f_{\pm}|^2$, a real function representing only the magnitude of the function $f_{\pm}$. Thus, introducing a frequency shift in one arm of the interferometer allows for the acquisition of additional phase information for biphotons \cite{Chen2021RPL,Fabre2022}. Since the N00N state interferometer ($\mu=0$ in Fig. \ref{Fig1}(b)) depends on the frequency sum for symmetric JSA, but the frequency difference for antisymmetric JSA \cite{JIN2018,Fabre2022,LiPRA2023}, it only allows to reconstruct the
amplitude of the JSA associated with the frequency sum for symmetric JSA or the frequency difference for antisymmetric JSA. However,with the generalized N00N state interferometer, it is possible to extract the phase information of the JSA due to the introduction of the frequency shift $\mu$ \cite{Chen2021RPL}. Specifically, one can measure the real part of $F_{+}$ associated with frequency sum for symmetric JSA (Eq.(\ref{Rsys})) and of $F_{-}$ associated with frequency difference for antisymmetric JSA (Eq.(\ref{Rant})), respectively. Then, if one introduces a phase $e^{i\pi/2}$ in one of the arms inside the generalized N00N state interferometer, the imaginary parts of $F_{\pm}$ can also be measured. Experimentally, such a phase can be realized with a quarter-wave plate. Once the real and imaginary parts are obtained, assuming that $f_{\pm}(0)\neq0$, the following reconstruction formula can be used:
\begin{equation}
\label{f+-}
f^*_{\pm}(\mu)=\frac{1}{2\pi f_{\pm}(0)}\int F_{\pm}(\mu,\tau)d\tau,
\end{equation}
to perform the full tomography of $f_+$ for symmetric JSA and of $f_-$ for antisymmetric JSA. This has been shown schematically in the right part of Fig.\ref{Fig1}(b). This reconstruction method is similar to the filter bank summation method in signal processing. The proof and an example for the reconstruction formula Eq.(\ref{f+-}) can be found in Appendix D. The analysis of the phase sensitivity for the generalized N00N state interferometer is similar to the case of the generalized HOM interferometer.  As an example, Fig. \ref{Fig3} show the simulated results of the real (left) and imaginary (right) parts of $F_+$ for a Gaussian input state (as shown in Appendix D) without a phase (a), and with a quadratic spectral phase (b)-(c). It can be seen that biphotons with different phases will lead to different distributions of the real (left) and imaginary (right) parts of $F_+$, which can be obtained from the coincidence probability measured with the generalized N00N state interferometer at different frequency shifts $\mu$. Conversely, the phase information of input state can also be extracted from $F_+$ using reconstruction formula Eq.(\ref{f+-}).

\section{\label{sec:4}Generalized combination interferometer}
The setup of a generalized combination interferometer is illustrated in Fig. \ref{Fig1}(c), where a frequency shift $\mu$ is introduced in the path 4. The coincidence probability between two detectors D5 and D6, as functions of the time delay $\tau$, $\tau_0$ and the frequency shift $\mu$, can be expressed as
\begin{eqnarray}
\label{R-C}
&&R(\tau_0,\tau,\mu)=\frac{1}{64}\int \int d\omega_s d\omega_i
\nonumber\\
&&\times \Big|f(\omega_s,\omega_i,\mu)(e^{-i\omega_s \tau_0}e^{-i(\omega_s+\mu)\tau}+e^{-i\omega_s \tau_0}+e^{-i(\omega_s+\mu) \tau}-1)(e^{-i\omega_i\tau_0}e^{-i(\omega_i+\mu)\tau}-e^{-i\omega_i \tau_0}-e^{-i(\omega_i+\mu) \tau}-1)\nonumber\\
&&+f(\omega_i,\omega_s,\mu)(e^{-i\omega_i \tau_0}e^{-i(\omega_i+\mu) \tau}+e^{-i(\omega_i+\mu) \tau}-e^{-i\omega_i \tau_0}+1)(e^{-i\omega_s \tau_0}e^{-i(\omega_s+\mu) \tau}-e^{-i(\omega_s+\mu) \tau}+e^{-i\omega_s \tau_0}+1)\Big|^2.
\end{eqnarray}
Also, the specific expression of Eq.(\ref{R-C}) depends on the exchange symmetry of the JSA. In general, $\tau_0$ is set to be fixed, and the coincidence probability in Eq.(\ref{R-C}) is a function of the time delay $\tau$ and the frequency shift $\mu$ \cite{LiPRA2023}. If the JSA is symmetric, i.e., $f(\omega_s, \omega_i)=f(\omega_i, \omega_s)$, we can simplify Eq.(\ref{R-C}) as (see Appendix E)
\begin{eqnarray}
\label{G-Rsys}
R_{S}(\tau_0,\tau,\mu)=&&\frac{1}{2}\Big(1-\frac{1}{2}Re[P_+(\tau_0)]Re[e^{-i\mu \tau}F_-(\mu,\tau)]-\frac{1}{2}Re[F_+(\mu,\tau)]-\frac{1}{2}Re[e^{-i\mu \tau}F_-(\mu,\tau)]\nonumber\\
&&+\frac{1}{4}Re[F_+(\mu,\tau+\tau_0)]+\frac{1}{4}Re[F_+(\mu,\tau-\tau_0)]\Big).
\end{eqnarray}
where $F_+$ and $F_-$ are the STFT of the function $f_+$ and $f_-$, respectively, as defined in Eq.(\ref{F+}) and Eq.(\ref{F-}). $P_+(\tau_0)$ is defined as
\begin{eqnarray}
\label{P+}
P_+(\tau_0)=\int d\omega_+f_+(\omega_+)f^*_+(\omega_+)e^{-i2\omega_+ \tau_0}.
\end{eqnarray}
Again, the additional phase factor $e^{-i\mu \tau}$ in Eq.(\ref{G-Rsys})  stems from the symmetric time-frequency
displacement operator $\hat{D}$ defined in Eq.(\ref{D}).
If we take $f_+$ as a Gaussian function, i.e., $f_+(\omega_+)=e^{-\omega_+^2/\sigma_+^2}e^{i\phi_+}$, where $\sigma_+$ denotes the linewidth of the pump pulse, then $Re[P_+(\tau_0)]\sim e^{-2\sigma_+^2\tau_0^2}$. The last two terms in Eq.(\ref{G-Rsys}) correspond to two identical interferograms  centered at $\pm\tau_0$. If $\tau_0<1/\sigma_+$, these two interferograms will gradually overlap and become fully indistinguishable as $\tau_0$ decreases to zero. Thus, if one would like to distinguish these two interferograms, $\tau_0$ must be much larger than the inverse linewidth $1/\sigma_+$ \cite{LiPRA2023}, resulting in $Re[P_+(\tau_0)] \sim 0$. Hence, we have
\begin{eqnarray}
\label{G-Rsys-1}
R_{S}(\tau_0,\tau,\mu)=&&\frac{1}{2}\Big(1-\frac{1}{2}Re[F_+(\mu,\tau)]-\frac{1}{2}Re[e^{-i\mu \tau}F_-(\mu,\tau)]\nonumber\\
&&+\frac{1}{4}Re[F_+(\mu,\tau+\tau_0)]+\frac{1}{4}Re[F_+(\mu,\tau-\tau_0)]\Big).
\end{eqnarray}
If the JSA is antisymmetric, i.e., $f(\omega_s, \omega_i)=-f(\omega_i, \omega_s)$, Eq.(\ref{R-C}) can be further simplified as
\begin{eqnarray}
\label{G-Rant}
R_{A}(\tau_0,\tau,\mu)=&&\frac{1}{2}\Big(1-\frac{1}{2}Re[P_-(\tau_0)]Re[F_+(\mu,\tau)]+\frac{1}{2}Re[F_+(\mu,\tau)]+\frac{1}{2}Re[e^{-i\mu \tau}F_-(\mu,\tau)]\nonumber\\
&&+\frac{1}{4}Re[e^{-i\mu \tau}F_-(\mu,\tau+\tau_0)]+\frac{1}{4}Re[e^{-i\mu \tau}F_-(\mu,\tau-\tau_0)]\Big).
\end{eqnarray}
where $P_-(\tau_0)$ has a similar definition as in Eq.(\ref{P+}). Analogously, to distinguish the two interferograms of the last two terms in Eq.(\ref{G-Rant}), $\tau_0$ must be much larger than the inverse linewidth $1/\sigma_-$, resulting in $Re[P_-(\tau_0)] \sim 0$. Thus, we have
\begin{eqnarray}
\label{G-Rant-1}
R_{A}(\tau_0,\tau,\mu)=&&\frac{1}{2}\Big(1+\frac{1}{2}Re[F_+(\mu,\tau)]+\frac{1}{2}Re[e^{-i\mu \tau}F_-(\mu,\tau)]\nonumber\\
&&+\frac{1}{4}Re[e^{-i\mu \tau}F_-(\mu,\tau+\tau_0)]+\frac{1}{4}Re[e^{-i\mu \tau}F_-(\mu,\tau-\tau_0)]\Big).
\end{eqnarray}
Note that since the general combination interferometer ($\mu=0$ in Fig.\ref{Fig1}(c)) depends on both the frequency sum and difference for symmetric JSA or antisymmetric JSA, it allows for the reconstruction of the amplitude of the JSA associated with both frequency sum and difference for symmetric JSA or antisymmetric JSA in a single quantum interferometer, as shown in our recent publication \cite{LiPRA2023}.  With the generalized combination interferometer ($\mu \neq0$ in Fig.\ref{Fig1}(c)), the interferograms associated with the real parts of $F_+$ and $F_-$ will appear in different regions of a single interferogram and can be well distinguished, as shown in Eq.(\ref{G-Rsys-1}) and Eq.(\ref{G-Rant-1}). On the other hand, the introduction of the frequency shift $\mu$ allows for the extraction of the phase information of the JSA, enabling the possibility of performing full tomography of biphotons. The method is similar to part III. First, one can measure simultaneously the real parts of $F_+$ and $F_-$ for symmetric JSA (Eq.(\ref{G-Rsys-1})) or antisymmetric JSA (Eq.(\ref{G-Rant-1})) with the setup of Fig.\ref{Fig1}(c). Then, by introducing a phase $e^{i\pi/2}$ in one of the arms inside the generalized combination interferometer, the imaginary parts of $F_+$ and $F_-$ can both be measured. Once the real and imaginary parts are obtained, assuming that $f_{\pm}(0)\neq0$, the reconstruction formulas defined in Eq.(\ref{f+-}) can be used to perform the full tomography of both $f_+$ and $f_-$ for symmetric JSA or antisymmetric JSA. This has been shown schematically in the right part of Fig.\ref{Fig1}(c).

\section{\label{sec:5}discussion}

By introducing a frequency shift in one arm of interferometers, we can obtain a generalized interferometer, which is sensitive to the phase of the input state. The interference between the frequency-shifted and frequency-unshifted paths at a 50/50 beam splitter allows for obtaining additional phase information of biphotons \cite{Chen2021RPL}. Consequently, it becomes possible to perform full tomography of biphotons, both in amplitude and phase, using such a generalized interferometer with the reconstruction formulas  Eq.(\ref{f-}) and Eq.(\ref{f+-}). From Eq.(\ref{R-HOM-W}), we can see that for the generalized HOM interferometer, the coincidence probability always depends on the frequency difference of biphotons associated with the phase matching part $f_-$,	​regardless of whether the JSA is symmetric, antisymmetric, or anyonic. As a result, it only allows for the reconstruction of the full complex JSA associated with the frequency difference for any symmetric JSA,  performing the partial tomography of biphotons.

From Eq.(\ref{Rsys}) and (\ref{Rant}), we can see that for the generalized N00N state interferometer, the coincidence probability depends on the frequency sum for the symmetric JSA and the frequency difference for the antisymmetric JSA. There exists a one-to-one correspondence between coincidence probability and  $F_+$ for symmetric JSA ( see Eq.(\ref{Rsys}) or $F_-$ for antisymmetric JSA (see Eq.(\ref{Rant})). This means that the real or imaginary part of $F_+$ or $F_-$ can be obtained directly from the coincidence data measured at different frequency shifts $\mu$, which enables the reconstruction of the full complex JSA associated with the frequency sum for symmetric JSA or the frequency difference for antisymmetric JSA, performing the partial tomography of biphotons as well.

For biphotons with anyonic symmetry, an interference arises between the symmetric and the antisymmetric part associated with the phase matching part $f_-$ in the generalized HOM interferometer, and an interference effect emerges between the symmetric part associated with $F_+$ and the antisymmetric one associated with $F_-$ in the generalized N00N state interferometer, due to the frequency shift $\mu$, as discussed in \cite{Fabre2022}. As a result, it is impossible to simultaneously perform full tomography of  both $f_+$ and $f_-$  with these two generalized interferometers.

However, in a generalized combination interferometer, the coincidence probability depends on both frequency sum and difference (see Eq.(\ref{G-Rsys-1}) and Eq.(\ref{G-Rant-1})) and has no one-to-one correspondence with $F_+$ or $F_-$ but rather with their combination. In this case, it is necessary to post-process the coincidence data to extract the desired information from $F_+$ and $F_-$. For example, in Eq.(\ref{G-Rsys-1}), one needs to first extract the real part of $F_+$ from the total coincidence data. Since $\tau_0$ must be much larger than the inverse linewidth $1/\sigma_+$, the coincidence data containing the last two terms in Eq.(\ref{G-Rsys-1}) are well distinguished from other parts. It is thus possible to extract the coincidence data associated with the real part of $F_+$ from the total coincidence data. Then, by substituting the real part of $F_+$ into Eq.(\ref{G-Rsys-1}), the real part of $F_-$ can be indirectly obtained. The imaginary parts of $F_+$ and $F_-$ can be obtained in a similar manner. With both the real and imaginary parts of $F_+$ and $F_-$, and using the reconstruction formulas defined in Eq.(\ref{f+-}), it becomes possible to perform the full tomography of both $f_+$ and $f_-$ for symmetric JSA. For antisymmetric JSA, the situation is analogous, where Eq.(\ref{G-Rant-1}) can be used to perform the full tomography of both $f_+$ and $f_-$. The main difference is the order of obtaining the real part of $F_-$ and $F_+$.

For biphotons with anyonic symmetry, the coincidence probability of the generalized combination interferometer becomes more complex, and Eq.(\ref{G-Rsys-1}) and Eq.(\ref{G-Rant-1}) do not hold. One possible solution to this issue is to create a superposition state  that satisfies the exchange symmetry condition, i.e., $\mathcal{F_S}(\omega_s,\omega_i)=f(\omega_s,\omega_i)+f(\omega_i,\omega_s)$, $\mathcal{F_A}(\omega_s,\omega_i)=f(\omega_s,\omega_i)-f(\omega_i,\omega_s)$, which are  symmetric and antisymmetric, respectively. This can be experimentally realized by placing a nonlinear crystal inside an interferometer, as reported in \cite{OE2018}. Then, the information about biphotons with $f_{\pm}$ can be extracted from $\mathcal{F_S}$ or $\mathcal{F_A}$. Thus, the proposed protocol for the full tomography of biphotons can also be extended to the case of biphotons with anyonic symmetry.

Experimentally, biphtons with symmetric, antisymmetric, or anyonic JSA can be generated in bulk nonlinear crystal or integrated optical devices using different methods \cite{PRL2012,OE2018,Optica20,ACS-P2021}. Additionally, the frequency shift can be experimentally realized using recent and promising electro-optics modulators \cite{Natrue2021,SR2021}, or a waveguide-based optomechanical system based on piezoelectric effect \cite{NP2016-1}. Another method for implementing such a frequency shift for wider spectral distributions involves introducing a dynamical shift of the eigenfrequencies or eigenmodes of optical resonators or waveguides \cite{NP2016}, which corresponds to a tuning of the photon frequency. In summary, it is feasible to experimentally perform the full tomography of biphotons using the generalized interferometers after carefully choosing biphoton resources, frequency shifts, and experimental setups.  If there is no frequency shift $\mu$ in generalized interferometers shown in Fig.(\ref{Fig1})($\mu=0$), they would revert to the original ones and only allow for the reconstruction of the amplitude of the corresponding $f$, namely $|f|$ \cite{LiPRA2023}. On the other hand, when the reconstruction formulas are used as a tool to analyze experimental data, typically in the form of a discretized coincidence distribution (i.e., an histogram) due to finite sampling, the reliability of the reconstruction results will be limited by imperfections such as losses and detector noise. The effect of such imperfections on the accuracy of the reconstruction formulas must be considered and needs to be studied further in the future.

\section{\label{sec:conclude}CONCLUSIONS}
By introducing a frequency shift in one arm of interferometers, we have  theoretically described three generalized quantum interferometers:the generalized HOM interferometer, the generalized N00N state interferometer, and the generalized combination interferometer. The key result is that these generalized interferometers are phase-sensitive to the input state. The interference between the frequency-shifted and frequency-unshifted paths at a 50/50 beam splitter enables the acquisition of additional phase information for biphotons within the generalized interferometers. Specifically, the generalized HOM interferometer allows the reconstruction of the full complex JSA associated with frequency difference for any symmetric JSA, while the N00N state interferometer allows the reconstruction of the full complex JSA associated with frequency sum for symmetric JSA or frequency difference for antisymmetric JSA. Both of them enable only partial tomography of biphotons, either in frequency difference or frequency sum. In contrast, the generalized combination interferometer enables the reconstruction of the full complex JSA associated with both the frequency sum and difference for symmetric JSA or antisymmetric JSA in a single interferometer, thereby allowing for the full tomography of biphotons. Furthermore, we have discussed the experimental feasibility and possible experimental difficulties with the proposed approach. This work provides an alternative method for the full characterization of an arbitrary two-photon state with exchange symmetry and holds potential for applications in high-dimensional quantum information processing.

\begin{acknowledgments}
This work has been supported by National Natural Science Foundation of China (12074309, 12074299, 12033007, 92365106, 61875205, 12103058, 61801458), the Youth Innovation Team of Shaanxi Universities, and the Natural Science Foundation
of Hubei Province (2022CFA039).
\end{acknowledgments}

\appendix

\section{The coincidence probability for a generalized HOM interferometer}

For a generalized HOM interferometer as described in Fig. \ref{Fig1}(a), the coincidence probability between two detectors D1 and D2, as functions of the time delay $\tau$ and the frequency shift $\mu$ , can be expressed as
\begin{equation}
\label{G-R-HOM}
R(\tau,\mu)=\frac{1}{4}\int \int d\omega_s d\omega_i\Big|f(\omega_i+\mu,\omega_s)e^{-i\omega_s \tau}-f(\omega_s+\mu,\omega_i)e^{-i\omega_i \tau}\Big|^2.
\end{equation}
For a normalized $f(\omega_s,\omega_i)$, using the relations,
\begin{eqnarray}
&&\omega_i+\mu-\omega_s=2[(\omega_i-\omega_s)/2+\mu/2]=2[\mu/2-\omega_-],\nonumber\\
&&\omega_s+\mu-\omega_i=2[(\omega_s-\omega_i)/2+\mu/2]=2[\mu/2+\omega_-],\nonumber\\
&&f(\omega_s,\omega_i)=f_+(\omega_+)f_-(\omega_-);f(\omega_i,\omega_s)=f_+(\omega_+)f_-(-\omega_-).
\end{eqnarray}
Eq. (\ref{G-R-HOM}) can be reduced to
\begin{eqnarray}
R(\tau,\mu)&&=\frac{1}{4}\int \int d\omega_+ d\omega_-\Big|f_+(\omega_+)f_-(\mu/2-\omega_-)e^{-i(\omega_+ +\omega_-)\tau}-f_+(\omega_+)f_-(\mu/2+\omega_-)e^{-i(\omega_+ -\omega_-)\tau}\Big|^2 \nonumber\\
&&=\frac{1}{4}\int d\omega_+\left|f_+(\omega_+)\right|^2 \int  d\omega_-\Big|f_-(\mu/2-\omega_-)e^{-i\omega_-\tau}-f_-(\mu/2+\omega_-)e^{i\omega_-\tau}\Big|^2\nonumber\\
&&=\frac{1}{4}\int  d\omega_-\Big|f_-(\mu/2-\omega_-)e^{-i\omega_-\tau}-f_-(\mu/2+\omega_-)e^{i\omega_-\tau}\Big|^2\nonumber\\
&&=\frac{1}{4}\int  d\omega_-\Big(f_-(\mu/2-\omega_-)e^{-i\omega_-\tau}-f_-(\mu/2+\omega_-)e^{i\omega_-\tau}\Big)\Big(f_-(\mu/2-\omega_-)e^{-i\omega_-\tau}-f_-(\mu/2+\omega_-)e^{i\omega_-\tau}\Big)^*\nonumber\\
&&=\frac{1}{4}\int  d\omega_-\Big(|f_-(\mu/2-\omega_-)|^2+|f_-(\mu/2+\omega_-)|^2\nonumber\\
&&-f_-(\mu/2-\omega_-)f_-^*(\mu/2+\omega_-)e^{-i2\omega_-\tau}
    -f_-^*(\mu/2-\omega_-)f_-(\mu/2+\omega_-)e^{i2\omega_-\tau}\Big)\nonumber\\
&&=\frac{1}{2}\Big(1-Re[W_-(\tau,\mu)]\Big).
\end{eqnarray}
where $W_-$ is the chronocyclic Wigner distribution of the phase matching function $f_-$ as defined in Eq.(\ref{W}). In above derivation, we change the variables in the double integral from $\omega_s,\omega_i$ to $\omega_+,\omega_-$, and use the normalized condition, i.e., $\int d\omega_+|f_+(\omega_+)|^2=\int d\omega_-|f_-(\mu/2-\omega_-)|^2=\int d\omega_-|f_-(\mu/2+\omega_-)|^2=1$.
\section{The proof and an example for the reconstruction formula Eq.(\ref{f-})}
Considering the integral $\int W_-(\tau,\mu/2)e^{i\mu \tau}d\tau$ with respect to $\tau$, and using Eq.(\ref{W}), we have
\begin{eqnarray}
\label{A4}
\int W_-(\tau,\mu/2)e^{i\mu \tau}d\tau &&=\int \int d\omega_-f_-(\mu/2-\omega_-)f_-^*(\mu/2+\omega_-)e^{-i2\omega_- \tau}e^{i\mu \tau}d\tau \nonumber\\
&&=\int d\omega_-f_-(\mu/2-\omega_-)f_-^*(\mu/2+\omega_-)\int e^{i(\mu-2\omega_-)\tau}d\tau\nonumber\\
&&=\int d\omega_-f_-(\mu/2-\omega_-)f_-^*(\mu/2+\omega_-)2\pi \delta(\mu-2\omega_-) \nonumber\\
&&=f_-(0)f_-^*(\mu)2\pi
\end{eqnarray}
we thus obtain Eq.(\ref{f-}).

As an example, we take the JSA associated with the phase-matching part $f_-$  as a Gaussian function,
\begin{equation}
\label{example1}
f_-(\omega_-)=e^{-\omega_-^2/\sigma_-^2}e^{i\phi_-},
\end{equation}
with the phase $\phi_-(\omega_-)=\alpha\omega_-^2$, a quadratic spectral phase with an index of $\alpha$. $\sigma_-$ denotes the linewidth determined by the phase-matching condition. Then, we get
\begin{eqnarray}
\label{f-f-}
f_-(\mu/2-\omega_-)=e^{-(\mu/2-\omega_-)^2/\sigma_-^2}e^{i[\alpha(\mu/2-\omega_-)^2]},\nonumber\\
f^*_-(\mu/2+\omega_-)=e^{-(\mu/2+\omega_-)^2/\sigma_-^2}e^{-i[\alpha(\mu/2+\omega_-)^2]}.
\end{eqnarray}
Substituting Eq.(\ref{f-f-}) into Eq.(\ref{W}), we obtain
\begin{eqnarray}
\label{W--}
W_-(\tau,\mu)=\frac{\sqrt{2\pi}\sigma_- }{2}e^{-\frac{\mu^2+(\alpha\mu+\tau)^2\sigma_-^4}{2\sigma_-^2}}.
\end{eqnarray}
Substituting Eq.(\ref{W--}) into Eq.(\ref{f-}), and using $f_-(0)=1$, we finally arrive at
\begin{equation}
\label{f--}
f^*_-(\mu)=\frac{1}{2}e^{-\mu^2/\sigma_-^2}e^{-i\alpha\mu^2}
\end{equation}
we thus recover the amplitude and phase as preset in Eq.(\ref{example1})  using the reconstruction formula Eq.(\ref{f-}).

\section{The coincidence probability for a generalized N00N state interferometer}
For a generalized N00N state interferometer as described in Fig. \ref{Fig1}(b), the coincidence probability between two detectors D3 and D4, as functions of the time delay $\tau$ and the frequency shift $\mu$,  can be expressed as
\begin{equation}
\label{R-NOON-1}
R(\tau,\mu)=\frac{1}{16}\int \int d\omega_s d\omega_i\Big|f(\omega_s,\omega_i,\mu)(e^{-i(\omega_s+\mu) \tau}+1)(e^{-i(\omega_i+\mu) \tau}+1)+f(\omega_i,\omega_s,\mu)(e^{-i(\omega_i+\mu) \tau}-1)(e^{-i(\omega_s+\mu) \tau}-1)\Big|^2.
\end{equation}
If the JSA is symmetric, i.e., $f(\omega_s, \omega_i)=f(\omega_i, \omega_s)$, we can further simplify Eq.(\ref{R-NOON-1}) as
\begin{eqnarray}
\label{R-sys}
R_{S}(\tau,\mu)&&=\frac{1}{2}\int \int d\omega_s d\omega_i\Big|f(\omega_s+\mu,\omega_i+\mu)+f(\omega_s,\omega_i)e^{i(\omega_s+\omega_i) \tau}\Big|^2\nonumber\\
&&=\frac{1}{2}\int \int d\omega_s d\omega_i\Big(f(\omega_s+\mu,\omega_i+\mu)+f(\omega_s,\omega_i)e^{i(\omega_s+\omega_i) \tau}\Big)\Big(f(\omega_s+\mu,\omega_i+\mu)+f(\omega_s,\omega_i)e^{i(\omega_s+\omega_i) \tau}\Big)^*\nonumber\\
&&=\frac{1}{2}\int \int d\omega_s d\omega_i\Big(|f(\omega_s+\mu,\omega_i+\mu)|^2+|f(\omega_s,\omega_i)|^2\nonumber\\
&&+f(\omega_s,\omega_i)f^*(\omega_s+\mu,\omega_i+\mu)e^{i(\omega_s+\omega_i) \tau}+f^*(\omega_s,\omega_i)f(\omega_s+\mu,\omega_i+\mu)e^{-i(\omega_s+\omega_i) \tau}\Big).
\end{eqnarray}
For a normalized $f(\omega_s,\omega_i)$, using the relations,
\begin{eqnarray}
\label{relation1}
&&\omega_s+\mu+(\omega_i+\mu)=2[(\omega_s+\omega_i)/2+\mu]=2(\omega_++\mu),\nonumber\\
&&\omega_s+\mu-(\omega_i+\mu)=2[(\omega_s-\omega_i)/2]=2\omega_-,\nonumber\\
&&f(\omega_s,\omega_i)=f_+(\omega_+)f_-(\omega_-);f(\omega_s+\mu,\omega_i+\mu)=f_+(\omega_++\mu)f_-(\omega_-).
\end{eqnarray}
Eq. (\ref{R-sys}) can be reduced to
\begin{eqnarray}
R_{S}(\tau,\mu)&&=\frac{1}{4}\int \int |f_-(\omega_-)|^2d\omega_-d\omega_+ \Big(|f_+(\omega_++\mu)|^2+|f_+(\omega_+)|^2\nonumber\\
&&+f_+(\omega_+)f^*(\omega_++\mu)e^{i2\omega_+ \tau}+f_+^*(\omega_+)f_+(\omega_++\mu)e^{-i2\omega_+ \tau}\Big)\nonumber\\
&&=\frac{1}{2}\Big(1+Re[F_+(\mu,\tau)]\Big).
\end{eqnarray}
where $F_+$ is the STFT of the function $f_+$ as defined in Eq.(\ref{F+}). In above derivation, we change the variables in the double integral from $\omega_s,\omega_i$ to $\omega_+,\omega_-$, and use the normalized condition, i.e., $\int d\omega_-|f_-(\omega_-)|^2=\int d\omega_+|f_+(\omega_++\mu)|^2=\int d\omega_+|f_+(\omega_+)|^2=1$.

If the JSA is antisymmetric, i.e., $f(\omega_s, \omega_i)=-f(\omega_i, \omega_s)$, we can further simplify Eq.(\ref{R-NOON-1}) as
\begin{eqnarray}
\label{R-ant}
R_{A}(\tau,\mu)&&=\frac{1}{2}\int \int d\omega_s d\omega_i\Big|f(\omega_s+\mu,\omega_i)e^{-i\omega_s \tau}+f(\omega_s,\omega_i+\mu)e^{-i\omega_i\tau}\Big|^2\nonumber\\
&&=\frac{1}{2}\int \int d\omega_s d\omega_i\Big(f(\omega_s+\mu,\omega_i)e^{-i\omega_s \tau}+f(\omega_s,\omega_i+\mu)e^{-i\omega_i\tau}\Big)\Big(f(\omega_s+\mu,\omega_i)e^{-i\omega_s \tau}+f(\omega_s,\omega_i+\mu)e^{-i\omega_i\tau}\Big)^*\nonumber\\
&&=\frac{1}{2}\int \int d\omega_s d\omega_i\Big(|f(\omega_s+\mu,\omega_i)|^2+|f(\omega_s,\omega_i+\mu)|^2\nonumber\\
&&+f(\omega_s+\mu,\omega_i)f^*(\omega_s,\omega_i+\mu)e^{-i(\omega_s-\omega_i) \tau}+f^*(\omega_s+\mu,\omega_i)f(\omega_s,\omega_i+\mu)e^{i(\omega_s-\omega_i) \tau}\Big).
\end{eqnarray}
For a normalized $f(\omega_s,\omega_i)$, using the relations,
\begin{eqnarray}
\label{relation2}
&&\omega_s-(\omega_i+\mu)=2[(\omega_s+\omega_i)/2-\mu/2]=2(\omega_--\mu/2),\nonumber\\
&&(\omega_s+\mu)-\omega_i=2[(\omega_s-\omega_i)/2+\mu/2]=2(\omega_-+\mu/2),\nonumber\\
&&f(\omega_s+\mu,\omega_i)=f_+(\omega_++\mu/2)f_-(\omega_-+\mu/2);\nonumber\\
&&f(\omega_s,\omega_i+\mu)=f_+(\omega_++\mu/2)f_-(\omega_--\mu/2).
\end{eqnarray}
Eq.(\ref{R-ant}) can be reduced to
\begin{eqnarray}
R_{A}(\tau,\mu)&&=\frac{1}{4}\int \int |f_+(\omega_++\mu/2)|^2d\omega_+d\omega_- \Big(|f_-(\omega_-+\mu/2)|^2+|f_-(\omega_--\mu/2)|^2\nonumber\\
&&+f_-(\omega_--\mu/2)f_-^*(\omega_-+\mu/2)e^{-i2\omega_- \tau}+f_-^*(\omega_--\mu/2)f_-(\omega_-+\mu/2)e^{i2\omega_- \tau}\Big)\nonumber\\
&&=\frac{1}{2}\Big(1+Re[e^{-i\mu \tau}F_-(\mu,\tau)]\Big).
\end{eqnarray}
where $F_-$ is the STFT of the function $f_-$ as defined in Eq.(\ref{F-}). In above derivation, we change the variables in the double integral from $\omega_s,\omega_i$ to $\omega_+,\omega_-$, and use the normalized condition, i.e., $\int d\omega_+|f_+(\omega_++\mu/2)|^2=\int d\omega_-|f_-(\omega_-+\mu/2)|^2=\int d\omega_-|f_-(\omega_--\mu/2)|^2=1$.
\section{The proof and an example for the reconstruction formula Eq.(\ref{f+-})}
Considering the integral $\int F_{\pm}(\mu,\tau)d\tau$ with respect to $\tau$, and using Eq.(\ref{F+}) and Eq.(\ref{F-}), we have
\begin{eqnarray}
\label{A4}
\int F_{\pm}(\mu,\tau)d\tau &&=\int \int d\omega_{\pm}f_{\pm}(\omega_{\pm})f_{\pm}^*(\omega_{\pm}+\mu)e^{-i2\omega_{\pm} \tau} d\tau \nonumber\\
&&=\int d\omega_{\pm}f_{\pm}(\omega_{\pm})f_{\pm}^*(\omega_{\pm}+\mu)\int e^{-i2\omega_{\pm} \tau} d\tau\nonumber\\
&&=\int d\omega_{\pm}f_{\pm}(\omega_{\pm})f_{\pm}^*(\omega_{\pm}+\mu)2\pi \delta(\omega_{\pm}) \nonumber\\
&&=f_{\pm}(0)f_{\pm}^*(\mu)2\pi
\end{eqnarray}
we thus obtain Eq.(\ref{f+-}).

As an example, we take the JSA associated with the energy-conservation part $f_+$  as a Gaussian function,
\begin{equation}
\label{example2}
f_+(\omega_+)=e^{-\omega_+^2/\sigma_+^2}e^{i\phi_+},
\end{equation}
with the phase $\phi_+(\omega_+)=\alpha\omega_+^2$, a quadratic spectral phase with an index of $\alpha$. $\sigma_+$ denotes the linewidth of the pump pulse. Then, we have
\begin{eqnarray}
\label{f+f+}
f_+(\omega_+)&&=e^{-\omega_+^2/\sigma_+^2}e^{i\alpha\omega_+^2},\nonumber\\
f^*_+(\omega_++\mu)&&=e^{-(\omega_++\mu)^2/\sigma_+^2}e^{-i[\alpha(\omega_++\mu)^2]}.
\end{eqnarray}
Substituting Eq.(\ref{f+f+}) into Eq.(\ref{F+}), we obtain
\begin{eqnarray}
\label{F++}
F_+(\mu,\tau)=\frac{\sqrt{2\pi}\sigma_+ }{2}e^{i\mu\tau}e^{-\frac{\mu^2+(\alpha\mu+\tau)^2\sigma_-^4}{2\sigma_-^2}}.
\end{eqnarray}
Substituting Eq.(\ref{F++}) into Eq.(\ref{f+-}), and using $f_+(0)=1$, we finally arrive at
\begin{equation}
\label{f--}
f^*_+(\mu)=\frac{1}{2}e^{-\mu^2/\sigma_+^2}e^{-i\alpha\mu^2}
\end{equation}
we thus recover the amplitude and phase as preset in Eq.(\ref{example2})  using the reconstruction formula Eq.(\ref{f+-}). $f^*_-(\mu)$ can also be reconstructed using Eq.(\ref{F-})  and Eq.(\ref{f+-}) in a similar manner.

\section{The coincidence probability for a generalized combination interferometer}
For a generalized combination interferometer as described in Fig. \ref{Fig1}(c), the coincidence probability between two detectors D5 and D6, as functions of the time delay $\tau_0$, $\tau$ and the frequency shift $\mu$, can be expressed as
\begin{eqnarray}
\label{G-R-com}
&&R(\tau_0,\tau,\mu)=\frac{1}{64}\int \int d\omega_s d\omega_i
\nonumber\\
&&\times \Big|f(\omega_s,\omega_i,\mu)(e^{-i\omega_s \tau_0}e^{-i(\omega_s+\mu)\tau}+e^{-i\omega_s \tau_0}+e^{-i(\omega_s+\mu) \tau}-1)(e^{-i\omega_i\tau_0}e^{-i(\omega_i+\mu)\tau}-e^{-i\omega_i \tau_0}-e^{-i(\omega_i+\mu) \tau}-1)\nonumber\\
&&+f(\omega_i,\omega_s,\mu)(e^{-i\omega_i \tau_0}e^{-i(\omega_i+\mu) \tau}+e^{-i(\omega_i+\mu) \tau}-e^{-i\omega_i \tau_0}+1)(e^{-i\omega_s \tau_0}e^{-i(\omega_s+\mu) \tau}-e^{-i(\omega_s+\mu) \tau}+e^{-i\omega_s \tau_0}+1)\Big|^2.
\end{eqnarray}
To simplify the expressions, we use $f_{(\omega_s, \omega_i)}$ to denote $f(\omega_s, \omega_i)$, , and similarly for others, in the following derivations. If the JSA is symmetric, i.e., $f_{(\omega_s, \omega_i)}=f_{(\omega_i, \omega_s)}$, we can further simplify Eq.(\ref{G-R-com}) as
\begin{eqnarray}
\label{G-R-com-S}
&&R_{S}(\tau_0,\tau,\mu)=\frac{1}{64}\int \int d\omega_s d\omega_i\nonumber\\
&&\times \Big|f_{(\omega_s,\omega_i)}-f_{(\omega_s+\mu,\omega_i)}e^{-i(\omega_s+\mu)\tau}+f_{(\omega_s,\omega_i+\mu)}e^{-i(\omega_i+\mu)\tau}-f_{(\omega_s,\omega_i)}e^{-i(\omega_s+\omega_i)\tau_0}-f_{(\omega_s+\mu,\omega_i)}e^{-i(\omega_s+\mu)\tau}e^{-i(\omega_s+\omega_i)\tau_0}\nonumber\\
&&-f_{(\omega_s+\mu,\omega_i+\mu)}e^{-i(\omega_s+\omega_i)\tau}+f_{(\omega_s,\omega_i+\mu)}e^{-i(\omega_i+\mu)\tau}e^{-i(\omega_s+\omega_i)\tau_0}+f_{(\omega_s+\mu,\omega_i+\mu)}e^{-i(\omega_s+\omega_i)\tau}e^{-i(\omega_s+\omega_i)\tau_0}\Big|^2\nonumber\\
&&=\frac{1}{64}\int \int d\omega_s d\omega_i\nonumber\\
&&\times \Big\{8\Big(|f_{(\omega_s,\omega_i)}|^2+|f_{(\omega_s+\mu,\omega_i)}|^2+|f_{(\omega_s,\omega_i+\mu)}|^2+|f_{(\omega_s+\mu,\omega_i+\mu)}|^2\Big)\nonumber\\
&&-4\Big(f_{(\omega_s+\mu,\omega_i)}f^*_{\omega_s,\omega_i+\mu}e^{-i(\omega_s-\omega_i)\tau}+f^*_{(\omega_s+\mu,\omega_i)}f_{(\omega_s,\omega_i+\mu)}e^{i(\omega_s-\omega_i)\tau}\Big)\nonumber\\
&&\times \Big(f_{(\omega_s,\omega_i)}f^*_{(\omega_s,\omega_i)}e^{-i(\omega_s+\omega_i)\tau_0}+f^*_{(\omega_s,\omega_i)}f_{(\omega_s,\omega_i)}e^{i(\omega_s+\omega_i)\tau_0}\Big)\nonumber\\
&&-8\Big(f_{(\omega_s+\mu,\omega_i+\mu)}f^*_{(\omega_s+\mu,\omega_i+\mu)}e^{-i(\omega_s+\omega_i)\tau}+f^*_{(\omega_s+\mu,\omega_i+\mu)}f_{(\omega_s+\mu,\omega_i+\mu)}e^{i(\omega_s+\omega_i)\tau}\Big)
\nonumber\\
&&-8\Big(f_{(\omega_s+\mu,\omega_i)}f^*_{(\omega_s,\omega_i+\mu)}e^{-i(\omega_s-\omega_i)\tau}+f^*_{(\omega_s+\mu,\omega_i)}f_{(\omega_s,\omega_i+\mu)}e^{i(\omega_s-\omega_i)\tau}\Big)\nonumber\\
&&+4\Big(f_{(\omega_s+\mu,\omega_i)}f^*_{(\omega_s,\omega_i+\mu)}e^{-i(\omega_s+\omega_i)\tau_0}e^{i(\omega_s+\omega_i)\tau}+f^*_{(\omega_s+\mu,\omega_i)}f_{(\omega_s,\omega_i+\mu)}e^{i(\omega_s+\omega_i)\tau_0}e^{-i(\omega_s+\omega_i)\tau}\Big)\nonumber\\
&&+4\Big(f_{(\omega_s+\mu,\omega_i)}f^*_{(\omega_s,\omega_i+\mu)}e^{-i(\omega_s+\omega_i)\tau_0}e^{-i(\omega_s+\omega_i)\tau}+f^*_{(\omega_s+\mu,\omega_i)}f_{(\omega_s,\omega_i+\mu)}e^{i(\omega_s+\omega_i)\tau_0}e^{i(\omega_s+\omega_i)\tau}\Big)\Big\}.
\end{eqnarray}
For a normalized $f(\omega_s,\omega_i)$, using Eq.(\ref{relation1}) and Eq.(\ref{relation2}), Eq.(\ref{G-R-com-S}) can be reduced to
\begin{eqnarray}
\label{G-R1}
&&R_{S}(\tau_0,\tau,\mu)=\frac{1}{2}\{1- \frac{1}{64}\int d\omega_+ \int d\omega_-\nonumber\\
&&+\Big\{4\Big(f_{(\omega_{-}+\mu/2,\omega_{-}-\mu/2)}f^*_{(\omega_{-}+\mu/2,\omega_{-}-\mu/2)}e^{-i2\omega_-\tau}+f^*_{(\omega_{-}+\mu/2,\omega_{-}-\mu/2)}f_{(\omega_{-}+\mu/2,\omega_{-}-\mu/2)}e^{i2\omega_-\tau}\Big)\nonumber\\
&&\times\Big(f_{\omega_+}f^*_{\omega_+}e^{-i2\omega_+\tau_0}+f^*_{\omega_+}f_{\omega_+}e^{i2\omega_+\tau_0}\Big)\nonumber\\
&&+8\Big(f_{(\omega_+,\omega_++\mu)}f^*_{(\omega_+,\omega_++\mu)}e^{-i2\omega_+\tau}+f^*_{(\omega_+,\omega_++\mu)}f_{(\omega_+,\omega_++\mu)}e^{i2\omega_+\tau}\Big)
\nonumber\\
&&+8\Big(f_{(\omega_-+\mu/2,\omega_--\mu/2)}f^*_{(\omega_-+\mu/2,\omega_--\mu/2)}e^{-i2\omega_-\tau}+f^*_{(\omega_-+\mu/2,\omega_--\mu/2)}f_{(\omega_-+\mu/2,\omega_--\mu/2)}e^{i2\omega_-\tau}\Big)\nonumber\\
&&-4\Big(f_{(\omega_+,\omega_++\mu)}f^*_{(\omega_+,\omega_++\mu)}e^{-i2\omega_+\tau_0}e^{-i2\omega_+\tau}+f^*_{(\omega_+,\omega_++\mu)}f_{(\omega_+,\omega_++\mu)}e^{i2\omega_+\tau_0}e^{i2\omega_+\tau}\Big)\nonumber\\
&&-4\Big(f_{(\omega_+,\omega_++\mu)}f^*_{(\omega_+,\omega_++\mu)}e^{-i2\omega_+\tau_0}e^{i2\omega_+\tau}+f^*_{(\omega_+,\omega_++\mu)}f_{(\omega_+,\omega_++\mu)}e^{i2\omega_+\tau_0}e^{-i2\omega_+\tau}\Big)\Big\}.
\end{eqnarray}
In above derivation, we change the variables in the double integral from $\omega_s,\omega_i$ to $\omega_+,\omega_-$, and use the normalized condition, i.e., $\int d\omega_-|f_-(\omega_-)|^2=\int d\omega_+|f_+(\omega_+)|^2=\int d\omega_+|f_+(\omega_++\mu/2)|^2=\int d\omega_+|f_+(\omega_+-\mu/2)|^2=\int d\omega_-|f_-(\omega_-+\mu/2)|^2=\int d\omega_-|f_-(\omega_--\mu/2)|^2=1$.
Finally, we arrive at
\begin{eqnarray}
R_{S}(\tau_0,\tau,\mu)=&&\frac{1}{2}\Big(1-\frac{1}{2}Re[P_+(\tau_0)]Re[e^{-i\mu \tau}F_-(\mu,\tau)]-\frac{1}{2}Re[F_+(\mu,\tau)]-\frac{1}{2}Re[e^{-i\mu \tau}F_-(\mu,\tau)]\nonumber\\
&&+\frac{1}{4}Re[F_+(\mu,\tau+\tau_0)]+\frac{1}{4}Re[F_+(\mu,\tau-\tau_0)]\Big).
\end{eqnarray}
where $F_+$ and $F_-$ are the STFT of the functions  $f_+$ and $f_-$, respectively, as defined in Eq.(\ref{F+}) and Eq.(\ref{F-}). If the JSA is antisymmetric, i.e., $f_{(\omega_s, \omega_i)}=-f_{(\omega_i, \omega_s)}$, the derivation process is similar and we give the result directly in Eq.(\ref{G-Rant}).

\end{CJK*}
\end{document}